\documentclass[aps,11pt]{revtex4}

\begin{document}

\tighten


\title{A Note on the Radiatively Induced Lorentz and CPT Violated Chern-Simons-like
Term in the Extended Quantum Electrodynamics}

\author{W.F. Chen}
\email{wengfenc@nipissingu.ca}
 \affiliation{Department of Mathematics, Nipissing University,
North Bay, Ontario, Canada P1B 8L7 \\
and\\
Department of Physics, University of Winnipeg, Winnipeg, Manitoba,
Canada R3B 2E9 }

\begin{abstract}
\noindent We show that the ambiguity for the Chern-Simons-like
term induced from quantum correction in the extended QED should
have nothing to do with the approximation on the exact fermionic
propagator, contradictory to the claim in Ref.\cite{paso}.
Further, we investigate the induced Chern-Simons-like term using
the original 't Hooft-Veltman dimensional regularization and
reproduce the result obtained by gauge symmetry analysis. This
fact demonstrates that the origin of the ambiguity should lie in
different choices on regularization
schemes.\\

\end{abstract}

\maketitle

\vspace{3ex}

It is more than tens years that the radiatively induced
Chern-Simons-like term in an extended QED has been keeping as a
popular topic. The most remarkable feature of this topic is the
ambiguity for the induced Chern-Simons-like term: a number of
distinct results have been obtained \cite{coleman}-- \cite{paso}.
As pointed out by Jackiw \cite{jackiw2}, this ambiguity is a
typical feature of quantum field theory when some fundamental
symmetry is violated that ``radiative corrections are finite but
undetermined". A convincing viewpoint is that the ambiguity
originates from the different regularization schemes of tackling
the linear UV divergence \cite{chen2}.
  However, recently it has been proposed  that the ambiguity may originate
  from different approximations
on the exact fermionic propagator and should be independent of the
utilization on the regularization methods \cite{paso}. In this
letter we show explicitly that the claimed difference in the
approximation on the exact fermionic propagator actually does not
exist. Further, we show that the origin for the discrepancy in the
radiatively induced Chern-Simons like term should attribute to the
choices on regularization schemes.

The fermionic sector of the extended quantum electrodynamics containing
a Lorentz- and CPT-violating
axial vector interaction with a constant four-vector  is
described by the following Lagrangian density:
\begin{eqnarray}
{\cal L}_{\rm F}=\overline{\psi} \left(i\partial\hspace{-2mm}/-e A\hspace{-2mm}/ -m
-\gamma_5 b \hspace{-1.6mm}/\right)\psi \ . \label{eq1}
\end{eqnarray}
The exact fermionic propagators with the constant vector field $b_\mu$ reads
\begin{eqnarray}
S(p,b)=\frac{i}{p\hspace{-1.6mm}/-m-b\hspace{-1.6mm}/  \gamma_5 }
\label{ferprop}
\end{eqnarray}

In the following we  show that the two different approximations on the
propagator (\ref{ferprop}) claimed in Ref.\,\cite{paso}
are actually identical. First, the approximation expansion of (\ref{ferprop})
 to the linear order of $b_\mu$
given in Ref.\,\cite{ezr} is
\begin{eqnarray}
S_{\rm I}(p,b)&=& \frac{i(p\hspace{-1.6mm}/ + m + b\hspace{-1.7mm}/  \gamma_5)}
{p^2-m^2+b^2-2 m  b\hspace{-1.6mm}/  \gamma_5+p\cdot b \gamma_5}
= \frac{i(p\hspace{-1.6mm}/ + m + b\hspace{-1.6mm}/  \gamma_5)}{p^2-m^2}
\left[ 1+\frac{2m b \hspace{-1.6mm}/-2p\cdot b\gamma_5}{p^2-m^2}\right]
+{\cal O}(b^2) \nonumber\\
&=&\frac{i(p\hspace{-1.6mm}/ + m + b\hspace{-1.6mm}/  \gamma_5)}{p^2-m^2}
-\frac{2i\gamma_5 (m b\hspace{-1.6mm}/ - p\cdot b) (p\hspace{-1.6mm}/ +m^2)}
{(p^2-m^2)^2}+{\cal O}(b^2)\nonumber\\
&=& \frac{i(p\hspace{-1.6mm}/ + m + b\hspace{-1.6mm}/  \gamma_5)}{p^2-m^2}
-\frac{2i\gamma_5 (m b\cdot p-p\hspace{-1.6mm}/ b\cdot p-m p\hspace{-1.6mm}/ b
\hspace{-1.6mm}/ +m^2 b\hspace{-1.6mm}/ )}{(p^2-m^2)^2}+{\cal O}(b^2)
 \label{approx1}
\end{eqnarray}

On the other hand, with the identity
\begin{eqnarray}
\frac{1}{A+B}=\frac{1}{A}-\frac{1}{A} B \frac{1}{A+B}=
\frac{1}{A}-\frac{1}{A} B\left( \frac{1}{A}-\frac{1}{A} B \frac{1}{A+B} \right),
\label{iden}
\end{eqnarray}
there arises  another approximation expansion of Eq.\,(\ref{ferprop}), which has been used in most of the literatures,
\begin{eqnarray}
 S_{\rm II}(p,b) &=& \frac{i}{p\hspace{-1.6mm}/ -m}+i \frac{1}{p\hspace{-1.6mm}/ -m} b\hspace{-1.6mm}/ \gamma_5 \frac{1}{p\hspace{-1.6mm}/ -m}+ {\cal O}(b^2) \nonumber\\
 &=&  \frac{i(p\hspace{-1.6mm}/ +m)}{p^2-m^2}+\frac{i (p\hspace{-1.6mm}/ +m) b\hspace{-1.6mm}/ \gamma_5 (p\hspace{-1.6mm}/ +m)}{(p^2-m^2)^2}\nonumber\\
 &=&\frac{i(p\hspace{-1.6mm}/ + m + b\hspace{-1.7mm}/  \gamma_5)}{p^2-m^2}
 +\frac{i\gamma_5\left(2p\cdot b p\hspace{-1.6mm}/-2 m^2 b\hspace{-1.7mm}/+m p\hspace{-1.6mm}/ b\hspace{-1.6mm}/ - m b\hspace{-1.6mm}/ p\hspace{-1.6mm}/ \right)}{(p^2-m^2)^2}+{\cal O}(b^2)\nonumber\\
 &=& \frac{i(p\hspace{-1.6mm}/ + m + b\hspace{-1.7mm}/  \gamma_5)}{p^2-m^2}
 -\frac{2i\gamma_5(m b\cdot p-p\hspace{-1.8mm}/ b\cdot p-m p\hspace{-1.8mm}/ b\hspace{-1.8mm}/ +m^2 b\hspace{-1.8mm}/ )}{(p^2-m^2)^2}+{\cal O}(b^2)
 \label{approx2}
\end{eqnarray}
One can see from Eqs.\,(\ref{approx1}) and (\ref{approx2}) that
 two approximation  expansions are identical: $S_{\rm I}(p,b)=S_{\rm II}(p,b)$.

 The above expansions are performed with no account taken on the
 regularization. To be more confirmative, we consider the
 the 't Hooft-Veltman dimensional regularization \cite{hooft,brma}
and  observe whether there exists difference between two approximation approaches
 at the regularization level.

The prescription for the 't Hooft-Veltman is defined as follows
\cite{hooft,brma}: the regularized $n$-dimensional space is
divided into  a direct sum of the original four-dimensional and
$n-4$-dimensional spaces. When one calculates the amplitude
represented
  by an Feynman diagram, the external momentum $p_\mu$  lives only in the
four-dimensional space, and the loop momentum $k_\mu$ in the whole $n$-dimensional
space,
\begin{eqnarray}
&& g_{\mu\nu}=\widetilde{g}_{\mu\nu}\oplus \widehat{g}_{\mu\nu}, ~~~
p_\mu = \widetilde{p}_\mu \oplus 0  ~~~
 k_\mu = \widetilde{k}_\mu \oplus \widehat{k}_\mu,
\nonumber\\
&& \widetilde{g}_{\mu\nu}\widehat{k}^\nu=\widehat{g}_{\mu\nu}\widetilde{k}^\nu=0,
~~\widetilde{g}_{\mu\nu}=\eta_{\mu\nu}, ~~~\widehat{g}_{\mu\nu}=-\delta_{\mu\nu}.
\label{presc1}
\end{eqnarray}
The $2^{[n/2]}\times 2^{[n/2]}$ $\gamma$-matrices are defined as
\begin{eqnarray}
&& \gamma_\mu = (\widetilde{\gamma}_\mu, \widehat{\gamma}_\mu),~~~
\gamma_5=\gamma^5\equiv i\widetilde{\gamma}^0\widetilde{\gamma}^1
\widetilde{\gamma}^2\widetilde{\gamma}^3
=\frac{i}{4!}\epsilon_{\mu\nu\lambda\rho}\widetilde{\gamma}^\mu
\widetilde{\gamma}^\nu\widetilde{\gamma}^\lambda\widetilde{\gamma}^\rho,\nonumber\\
&& \left\{\gamma_\mu, \gamma_\nu\right\}=2 g_{\mu\nu},
~~~\left\{\widetilde{\gamma}_\mu, \widetilde{\gamma}_\nu\right\}
=2 \widetilde{g}_{\mu\nu}, ~~~\left\{\widehat{\gamma}_\mu, \widehat{\gamma}_\nu\right\}
=2 \widehat{g}_{\mu\nu}, \nonumber\\
&& \{\widetilde{\gamma}_\mu, \widehat{\gamma}_\nu\}=0,
~~\{\gamma_5,\widetilde{\gamma}_\mu\}=0, ~~~[\gamma_5, \widehat{\gamma}_\mu]=0,
\label{presc2}
\end{eqnarray}
where $\widetilde{\gamma}_\mu$ has only non-vanishing $4\times 4$ components in the
upper-left corner  and
$\widehat{\gamma}_\mu$ has only the non-vanishing $2^{-\epsilon}\times 2^{-\epsilon}$
elements in the lower-right corner, $\epsilon\equiv 2-n/2$. In addition, like the
external momentum,  the $\gamma$-matrix appearing in the external vertex lives
in four dimensions.  One can easily see that the above regularization prescription
presents the $SO(1,3) \times SO(n-3)$ covariance rather than the whole $SO(n)$ covariance.

According to the above prescription, the first  approximation
expansion on the exact fermionic propagator appearing in a loop of
a certain Feynman diagram with a loop momentum $k$ is
 \begin{eqnarray}
S^R_{\rm I}(k,b)&=&\frac{i}{\widetilde{k\hspace{-2mm}/} + \widehat{k\hspace{-2mm}/}
-m-b\hspace{-1.8mm}/ \gamma_5}
=\frac{i\left(\widetilde{k\hspace{-2mm}/ } + \widehat{k\hspace{-2mm}/}
+m+b\hspace{-1.8mm}/ \gamma_5\right)}{
\widetilde{k}^2-\widehat{k}^2-m^2+b^2-2 m b\hspace{-1.8mm}/ \gamma_5
+2 \widetilde{k}\cdot {b}\gamma_5-2 b\hspace{-1.8mm}/ \widehat{k\hspace{-2mm}/}\gamma_5}
\nonumber\\
&=& \frac{i\left(\widetilde{k\hspace{-2mm}/ } + \widehat{k\hspace{-2mm}/ }
+m+b\hspace{-1.8mm}/ \gamma_5\right)}{\widetilde{k}^2-\widehat{k}^2-m^2}
-\frac{2i\gamma_5 \left(m  b\hspace{-1.8mm}/-\widetilde{k}\cdot {b} \right)
\left(\widetilde{k\hspace{-2mm}/ } + \widehat{k\hspace{-2mm}/}
+m \right)} {\left(\widetilde{k}^2-\widehat{k}^2-m^2\right)^2}\nonumber\\
&& +\frac{2i\gamma_5 b\hspace{-1.8mm}/ \left(m \widehat{k\hspace{-2mm}/}- \widehat{k}^2-
\widetilde{k\hspace{-2mm}/} \widehat{k\hspace{-2mm}/ }
\right)}{\left(\widetilde{k}^2-\widehat{k}^2-m^2\right)^2}+{\cal O}(b^2)
\label{differ1}
\end{eqnarray}
While the second approximation leads to
\begin{eqnarray}
S^R_{\rm II}(k,b)&=& \frac{i\left(\widetilde{k\hspace{-2mm}/ }
 + \widehat{k\hspace{-2mm}/ }
+m\right)}{\widetilde{k}^2-\widehat{k}^2-m^2}
+\frac{i\left(\widetilde{k\hspace{-2mm}/ } + \widehat{k\hspace{-2mm}/ }
+m\right)b\hspace{-1.8mm}/ \gamma_5 \left(\widetilde{k\hspace{-2mm}/ }
+ \widehat{k\hspace{-2mm}/ }
+m\right)}{\left(\widetilde{k}^2-\widehat{k}^2-m^2\right)^2}+{\cal O}(b^2)\nonumber\\
&=& \frac{i\left(\widetilde{k\hspace{-2mm}/ } + \widehat{k\hspace{-2mm}/ }
+m+b\hspace{-1.8mm}/ \gamma_5\right)}{\widetilde{k}^2-\widehat{k}^2-m^2}
-\frac{2i\gamma_5 \left(m  b\hspace{-1.8mm}/-\widetilde{k}\cdot {b} \right)
\left(\widetilde{k\hspace{-2mm}/ } + \widehat{k\hspace{-2mm}/}
+m \right)} {\left(\widetilde{k}^2-\widehat{k}^2-m^2\right)^2}\nonumber\\
&& +\frac{2i\gamma_5 b\hspace{-1.8mm}/ \left(m \widehat{k\hspace{-2mm}/}-
\widehat{k}^2-
\widetilde{k\hspace{-2mm}/} \widehat{k\hspace{-2mm}/ }
\right)}{\left(\widetilde{k}^2-\widehat{k}^2-m^2\right)^2}+{\cal O}(b^2)
\label{differ2}
\end{eqnarray}
Eqs.\,(\ref{differ1}) and (\ref{differ2}) show that even at the 't
Hooft-Veltman dimensional regularization level, the two $b$-linear
order approximations for the exact fermionic propagator appearing
in the fermionic loop are still identical:
\begin{eqnarray}
S^R_{\rm I}(k,b)=S^R_{\rm II}(k,b)\equiv S^R (k,b)
\end{eqnarray}

In the following we observe the induced Chern-Simons-like terms
using the prescription defined in (\ref{presc1}) and
(\ref{presc2}). The one-loop vacuum polarization tensor in the 't
Hooft-Veltman dimensional regularization is
\begin{eqnarray}
i\Pi_{\mu\nu}^{(1)R}(p,b)&=&-\int \frac{d^nk}{(2\pi)^n}\mbox{Tr}\left[
(-ie\mu^{\epsilon}\gamma_\mu)S^R(k)(-ie\mu^{\epsilon}\gamma_\nu)S^R(k+p) \right]
\nonumber\\
&=&-e^2\mu^{2\epsilon}\int \frac{d^n}{(2\pi)^n}\mbox{Tr}
\left(\gamma_\mu\frac{1}{ k\hspace{-2mm}/ -m- b\hspace{-1.6mm}/ \gamma_5}
\gamma_\nu \frac{1}{k\hspace{-2mm}/
+p\hspace{-1.8mm}/- b\hspace{-1.6mm}/ \gamma_5 }\right)
\end{eqnarray}
Using the expansion (\ref{iden}), we  focus on the sector of
the vacuum polarization tensor
relevant to the Chern-Simons-like term, which consists of the $b$-linear terms
in $\Pi_{\mu\nu}^{(1)}(p,b)$,
\begin{eqnarray}
i\Pi_{\mu\nu}^{\rm CS}(p,b)&=&
i\Pi_{\mu\nu}^{(1) {\rm CS}}(p,b)+i\Pi_{\mu\nu}^{(2) {\rm CS}}(p,b),\nonumber\\
i\Pi_{\mu\nu}^{(1){\rm CS}}(p,b) &=& -e^2 \mu^{2\epsilon}\int
\frac{d^nk}{(2\pi)^n}\mbox{Tr}\left(\gamma_\mu\frac{1}{ k\hspace{-2mm}/ -m}
 b\hspace{-1.6mm}/ \gamma_5 \frac{1}{ k\hspace{-2mm}/ -m}
 \gamma_\nu \frac{1}{k\hspace{-2mm}/
+p\hspace{-1.6mm}/ -m } \right), \nonumber\\
i\Pi_{\mu\nu}^{(2){\rm CS}}(p,b) &=& -e^2 \mu^{2\epsilon}\int
\frac{d^nk}{(2\pi)^n}\mbox{Tr}\left(\gamma_\mu\frac{1}{ k\hspace{-2mm}/ -m}
 \gamma_\nu \frac{1}{ k\hspace{-2mm}/ +p\hspace{-1.6mm}/-m}
 b\hspace{-1.6mm}/ \gamma_5 \frac{1}{k\hspace{-2mm}/
+p\hspace{-1.6mm}/ -m } \right).
\end{eqnarray}

Further, one can (naively) show that $\Pi_{\mu\nu}^{(1){\rm CS}}(p,b)$
and $\Pi_{\mu\nu}^{(2){\rm CS}}(p,b)$
are identical. Making the shift $k\to k-p$ in $\Pi_{\mu\nu}^{(2){\rm CS}}(p,b)$,
we have
\begin{eqnarray}
i\Pi_{\mu\nu}^{(2){\rm CS}}(p,b) &=& -e^2 \mu^{2\epsilon}\int
\frac{d^nk}{(2\pi)^n}\mbox{Tr}\left(\gamma_\mu
\frac{1}{ k\hspace{-2mm}/ -p\hspace{-1.6mm}/-m}
 \gamma_\nu \frac{1}{ k\hspace{-2mm}/-m}
 b\hspace{-1.6mm}/ \gamma_5 \frac{1}{k\hspace{-2mm}/
 -m } \right)\nonumber\\
 &=&  -e^2 \mu^{2\epsilon}\int
\frac{d^nk}{(2\pi)^n}\mbox{Tr}\left(\gamma_\nu \frac{1}{ k\hspace{-2mm}/-m}
 b\hspace{-1.6mm}/ \gamma_5 \frac{1}{k\hspace{-2mm}/
 -m } \gamma_\mu \frac{1}{ k\hspace{-2mm}/ -p\hspace{-1.6mm}/-m}\right)
 \nonumber\\
 &=& i\Pi_{\nu\mu}^{(1) {\rm CS}}(-p,b)=
 \Pi_{\mu\nu}^{(1) {\rm CS}}(p,b)
 \label{shift}
\end{eqnarray}
where in the last line we have used the tensor structure of
$\Pi_{\mu\nu}^{\rm CS}(p,b)=
\epsilon_{\mu\nu\lambda\rho} p^\lambda b^\rho f(p^2)$. Therefore, we obtain
\begin{eqnarray}
&& i\Pi_{\mu\nu}^{\rm CS}(p,b)= 2i\Pi_{\mu\nu}^{(1){\rm CS}}(p,b)\nonumber\\
&=& -2e^2 \mu^{2\epsilon}\int
\frac{d^nk}{(2\pi)^n}\frac{\mbox{Tr}\left[\gamma_\mu
\left(\widetilde{k\hspace{-2mm}/} +\widehat{k\hspace{-2mm}/}+m \right)
b\hspace{-1.8mm}/ \gamma_5 \left(\widetilde{k\hspace{-2mm}/}
 +\widehat{k\hspace{-2mm}/}+m \right)\gamma_\nu \left(\widetilde{k\hspace{-2mm}/}
  + {p\hspace{-1.8mm}/} +\widehat{k\hspace{-2mm}/}+m \right)\right]}{(k^2-m^2)^2
  [(k+p)^2-m^2]}\nonumber\\
&=&8i e^2\mu^{2\epsilon}\epsilon_{\mu\nu\lambda\rho}
\int \frac{d^4\widetilde{k}}{(2\pi)^4}\int\frac{d^{n-4}\widehat{k}}{(2\pi)^{n-4}}
\frac{2\widetilde{k}\cdot b \widetilde{k}^\lambda p^\rho
+(\widetilde{k}^2-\widehat{k}^2 -m^2)\widetilde{k}^\lambda b^\rho
+ (\widetilde{k}^2+\widehat{k}^2 +m^2){p}^\lambda b^\rho}
{(\widetilde{k}^2-\widehat{k}^2-m^2)^2 [(\widetilde{k}+p)^2-\widehat{k}^2-m^2]}\nonumber\\
\label{cstens}
\end{eqnarray}
In above  we have used the prescriptions listed in (\ref{presc1}) and (\ref{presc2})
and the following formula,
\begin{eqnarray}
\mbox{Tr}\left(\gamma_5\widetilde{\gamma}_\mu
\widetilde{\gamma}_\nu \widetilde{\gamma}_\lambda
\widetilde{\gamma}_\rho\right)=-4i \epsilon_{\mu\nu\lambda\rho},
~~~~\widehat{k\hspace{-2mm}/}\widehat{k\hspace{-2mm}/}=-~\widehat{k}^2
\end{eqnarray}

To evaluate the non-covariant integrals in  Eq.\,(\ref{cstens}), it is very convenient
to repeatedly use the following decomposition to improve the UV behavior of the
integrand,
\begin{eqnarray}
\frac{1}{(k+p)^2-m^2}=\frac{1}{k^2-m^2}-\frac{2k\cdot p+p^2}{(k^2-m^2)[(k+p)^2-m^2]}.
\end{eqnarray}
Once the integration becomes UV finite, one can directly take the
$n\to 4$ limit before performing the integration, and the
integrand proportional to the evanescent quantity
$\widehat{k}_\mu$ will vanish automatically \cite{martin}. For
examples, we have
\begin{eqnarray}
&&\lim_{n \to 4}\int \frac{d^nk}{(2\pi)^n}\frac{\widehat{k}^2}{(k^2-m^2)^2 [(k+p)^2-m^2]}
\nonumber\\
&=&\lim_{n \to 4}\int \frac{d^nk}{(2\pi)^n}\frac{\widehat{k}^2}{(k^2-m^2)^2}\left(
\frac{1}{k^2-m^2}-\frac{2k\cdot p+p^2}{(k^2-m^2)[(k+p)^2-m^2]} \right)\nonumber\\
&=&\lim_{n \to 4}
\frac{1}{(2\pi)^n}\int d^4 \widetilde{k}\int d^{n-4}\widehat{k}\frac{\widehat{k}^2}
{(\widetilde{k}^2-\widehat{k}^2-m^2)^3}
= \frac{i}{32\pi^2};
\end{eqnarray}
\begin{eqnarray}
&&\lim_{n \to 4}\int \frac{d^nk}{(2\pi)^n}\frac{\widetilde{k}_\mu \widetilde{k}_\nu}
{(k^2-m^2)^2 [(k+p)^2-m^2]}
\nonumber\\
&=&\lim_{n \to 4}\int \frac{d^nk}{(2\pi)^n}\frac{\widetilde{k}_\mu \widetilde{k}_\nu}
{(k^2-m^2)^2} \left( \frac{1}{k^2-m^2}
-\frac{2 \widetilde{k}\cdot p+p^2}{(k^2-m^2)[(k+p)^2-m^2]}\right)
\nonumber\\
&=&\frac{1}{4}\widetilde{g}_{\mu\nu}\lim_{n \to 4}
\int \frac{d^4\widetilde{k}}{(2\pi)^4}\widetilde{k}^2
\int \frac{d^{n-4}\widehat{k}}{(2\pi)^{n-4}}
\frac{1}{(\widetilde{k}^2-\widehat{k}^2-m^2)^3}
-\int \frac{d^4k}{(2\pi)^4}\frac{ (2 k\cdot p+p^2) k_\mu k_\nu}{(k^2-m^2)^3
[(k+p)^2-m^2]}\nonumber\\
&=&\frac{i}{64\pi^2}g_{\mu\nu}\left[\frac{1}{\epsilon}
-\gamma+\ln \frac{4\pi\mu^2}{m^2}+
2-\frac{2m}{p}
\left( 4-\frac{p^2}{m^2}\right)^{1/2}
\arctan \frac{p/m}{( 4-{p^2}/{m^2})^{1/2}}\right]\nonumber\\
&&+ \frac{i}{32\pi^2}\frac{p_\mu p_\nu}{p^2}\left[
1-4\frac{m}{p}\frac{1}{(4-p^2/m^2)^{1/2}}\arctan \frac{p/m}{( 4-p^2/m^2)^{1/2}}
\right]
\end{eqnarray}
Note that in above integrations one should integrate over
$\widehat{k}$  first and then over $\widetilde{k}$.

Using the integration formula listed in Appendix,  we can obtain the part
of the vacuum polarization tensor relevant to the Chern-Simon-like term,
\begin{eqnarray}
i\Pi_{\mu\nu}^{\rm CS}(p,b)&=&\frac{ie^2}{2\pi^2}  \epsilon_{\mu\nu\lambda\rho}
p^\lambda b^\rho
\left[\frac{1}{2}-\int_0^1 dx \frac{1-x}{1-x(1-x)p^2/m^2}\right.\nonumber\\
&&\left.+\frac{p^2}{m^2}\int_0^1 dx \frac{5 x(1-x)^2-2 (1-x)^3-x (1-x)}
{1-x(1-x)p^2/m^2}\right]\nonumber\\
&=&\frac{ie^2}{2\pi^2}  \epsilon_{\mu\nu\lambda\rho}p^\lambda b^\rho
\left[-4+\frac{4m}{p} \left(4-\frac{p^2}{m^2}\right)^{1/2}
\arctan \frac{p/m}{(4-p^2/m^2)^{1/2}}\right]
\label{result1}
\end{eqnarray}
The induced Chern-Simons coefficient $k_\mu$ is defined from
$\Pi_{\mu\nu}^{\rm CS}(p,b)$ at $p^2=0$. Eq.\,(\ref{result1}) gives
\begin{eqnarray}
k_\mu=0
\label{vanish}
\end{eqnarray}

However, this is not the end of story, one should evaluate the
surface term produced by the loop momentum shift $k\to k-p$, since
the loop integral in linearly divergent.
 As shown in Ref.\,\cite{jackiw}, the shift can somehow lead to a
non-vanishing  surface term.
 In spite of the overwhelming viewpoint that the shift can be safely
 taken after having implemented a regularization, but due to the
 particularity of the t'Hooft-Veltman regularization scheme,
 an explicit calculation should be performed to observe whether
 a finite result can arise from the surface term. The calculation
 on the chiral anomaly from the triangle diagram is a typical
 example \cite{bert}.

The surface term induced by the variable shift in Eq.\,(\ref{shift}) is
\begin{eqnarray}
\Delta\Pi_{\mu\nu}^{(2) CS}(p,b) &=& ie^2 \mu^{2\epsilon}\int
\frac{d^nk}{(2\pi)^n}\left[\mbox{Tr}\left(\gamma_\mu\frac{1}{ k\hspace{-2mm}/ -m}
 \gamma_\nu \frac{1}{ k\hspace{-2mm}/ +p\hspace{-1.6mm}/-m}
 b\hspace{-1.6mm}/ \gamma_5 \frac{1}{k\hspace{-2mm}/
+p\hspace{-1.6mm}/ -m } \right)\right.\nonumber\\
&&\left.-\mbox{Tr}\left(\gamma_\mu\frac{1}{ k\hspace{-2mm}/ -p\hspace{-1.6mm}/-m}
 \gamma_\nu \frac{1}{ k\hspace{-2mm}/-m}
 b\hspace{-1.6mm}/ \gamma_5 \frac{1}{k\hspace{-2mm}/
 -m } \right)\right]
 \label{surface}
\end{eqnarray}
The calculation on the surface term is the same as that evaluating
the chiral anomaly from the triangle diagram $\langle j_\mu
(p)j_\nu (p-q) j_\rho^5(q)\rangle$  with the momentum of the axial
vector current  $q=0$. In the 't Hooft-Veltman dimensional
regularization, the surface term can only get contribution from
the integrands containing the evanescent momentum $\widehat{k}^2$
\cite{bert}. Therefore, evaluating the trace  in (\ref{surface})
and taking into account the terms containing $\widehat{k}^2$, we
have
\begin{eqnarray}
\Delta\Pi_{\mu\nu}^{(2) CS}(p,b) &=& 4e^2
\epsilon_{\mu\nu\lambda\rho}\int
\frac{d^nk}{(2\pi)^n}\left[\frac{\widehat{k}^2
\left(\widetilde{k}^\lambda+2p^{\lambda}
\right)b^\rho}{(k^2-m^2)[(k+p)^2-m^2]^2}
-\frac{\widehat{k}^2\left(\widetilde{k}^\lambda+p^{\lambda}
\right)b^\rho}
{(k^2-m^2)^2[(k-p)^2-m^2]}\right]\nonumber\\
&=&\frac{ie^2}{24\pi^2} \epsilon_{\mu\nu\lambda\rho}p^\lambda
b^\rho \left(-2+6-1-3\right)=0.\label{surf2}
\end{eqnarray}
This coincides with the result obtained in Ref.\,\cite{bonn}.

In summary, we have shown the two approximations on the exact
fermionic propagator used in Ref.\,\cite{paso} are actually
identical at both unregularized and regularization level. The
discrepancy shown by the author should come from the different
algebraic operations the author made on the integrand rather than
taking approximation on the fermionic propagator. Further, using
the 't Hooft-Veltman dimensional regularization, we have
calculated the induced Chern-Simons-like term   and obtain a
vanishing result. The reason for this result is probably because
the dimensional regulation preserves the Ward identities and make
the theory well defined at regularization level, as analyzed in
Ref.\,\cite{bonn}. In comparison with the various results obtained
in the literature \cite{coleman}-- \cite{paso},  the origin for
the ambiguity in the induced Chern-Simons terms should lie in
different choices on regularization schemes.

\vspace{6mm}

 \noindent \textbf{Acknowledgement:} This work is supported by the Natural Sciences and
 Engineering Research Council of Canada. I would like to thank Professor R. Jackiw for
 encouragement. I also would like to thank Professor G. Bonneau for continuous
comments.

 \appendix

 \section{Integration formula}

 Some integration formula used to evaluate (\ref{cstens}) and (\ref{surface})
 are listed in this appendix:
\begin{eqnarray}
&&\lim_{n\to
4}\int\frac{d^nk}{(2\pi)^n}\frac{\widehat{k}^2}{(k^2-m^2)^2[(k+p)^2-m^2]}
=\lim_{n\to
4}\int\frac{d^nk}{(2\pi)^n}\frac{\widehat{k}^2}{(k^2-m^2)[(k+p)^2-m^2]^2}
\nonumber\\
&&=\lim_{n\to
4}\int\frac{d^nk}{(2\pi)^n}\frac{\widehat{k}^2}{(k^2-m^2)^3}=
\lim_{n\to 4}\frac{1}{(2\pi)^n}\int d^4\widetilde{k}\int
d^{n-4}\widehat{k}\frac{\widehat{k}^2}{(\widetilde{k}^2-\widehat{k}^2-m^2)^3}
=\frac{i}{32\pi^2},\\
&&\lim_{n\to
4}\int\frac{d^nk}{(2\pi)^n}\frac{\widehat{k}^2\widetilde{k}_\mu}{(k^2-m^2)^2[(k-p)^2-m^2]}
=\int\frac{d^nk}{(2\pi)^n}\frac{2\widehat{k}^2\widetilde{k}\cdot
p\widetilde{k}_\mu}
{(k^2-m^2)^4}\nonumber\\
 && = \lim_{n\to 4}\frac{1}{(2\pi)^n}\int
d^4\widetilde{k}\int
d^{n-4}\widehat{k}\frac{(-2)\widehat{k}^2\widetilde{k}_\mu}
{(\widetilde{k}^2-\widehat{k}^2-m^2)^4}
=\frac{i}{96\pi^2} p_\mu,\\
&&\lim_{n\to 4}\int\frac{d^nk}{(2\pi)^n}
\frac{\widehat{k}^2\widetilde{k}_\mu}{(k^2-m^2)[(k+p)^2-m^2]^2}
=\lim_{n\to
4}\int\frac{d^nk}{(2\pi)^n}\frac{(-4)\widetilde{k}_\mu\widehat{k}^2
\widetilde{k}\cdot p}{(k^2-m^2)^3}=-\frac{i}{48\pi^2}p_\mu\\
 &&
\int\frac{d^4k}{(2\pi)^4}\frac{1}{(k^2-m^2)^2[(k+p)^2-m^2]}
= -\frac{i}{16\pi^2}\int_0^1 dx\frac{1-x}{1-x(1-x)p^2/m^2},\\
&& \int\frac{d^4k}{(2\pi)^4}\frac{\widetilde{k}_\mu}{(k^2-m^2)^2[(k+p)^2-m^2]}
= \frac{i}{16\pi^2}p_\mu \frac{1}{m^2}\int_0^1 dx\frac{x(1-x)}{1-x(1-x)p^2/m^2},\\
&&\lim_{n\to 4}\int\frac{d^nk}{(2\pi)^n}\frac{\widetilde{k}_\mu\widetilde{k}_\nu}
{(k^2-m^2)^2[(k+p)^2-m^2]}
= \frac{i}{64\pi^2}{g}_{\mu\nu}\left(\frac{1}{\epsilon}-\gamma +
\ln\frac{4\pi\mu^2}{m^2} \right)\nonumber\\
&& +\frac{i}{64\pi^2}{g}_{\mu\nu}\frac{p^2}{m^2}\int_0^1dx
\frac{(1-x)^3-x (1-x)^2}{1-x(1-x)p^2/m^2}-\frac{i}{16\pi^2}\frac{p_\mu p_\nu}{p^2}
\frac{p^2}{m^2}\int_0^1 dx\frac{x(1-x)^2}{1-x(1-x)p^2/m^2},\\
&& \lim_{n\to 4}\int\frac{d^nk}{(2\pi)^n}
\frac{\widetilde{k}^2\widetilde{k}_\mu}{(k^2-m^2)^2[(k+p)^2-m^2]}
= \lim_{n\to 4} \int\frac{d^nk}{(2\pi)^n}
\frac{\widetilde{k}_\mu}{(k^2-m^2)[(k+p)^2-m^2]}\nonumber\\
&&+\lim_{n\to 4}\int\frac{d^nk}{(2\pi)^n}
\frac{\widetilde{k}_\mu \widehat{k}^2}{(k^2-m^2)^2[(k+p)^2-m^2]}
+m^2\int\frac{d^4k}{(2\pi)^4}
\frac{\widetilde{k}_\mu }{(k^2-m^2)^2[(k+p)^2-m^2]}
\nonumber\\
&&=-\frac{i}{32\pi^2}p_\mu \left(\frac{1}{\epsilon}-\gamma +
\ln\frac{4\pi\mu^2}{m^2} \right)-\frac{i}{32\pi^2}p_\mu \frac{p^2}{m^2}
\int_0^1 dx\frac{(1-x)^3-5x(1-x)^2+2 x (1-x)}{1-x(1-x)p^2/m^2}\nonumber\\
&& -\frac{i}{96\pi^2}p_\mu +\frac{i}{16\pi^2}p_\mu
\int_0^1 dx\frac{ x (1-x)}{1-x(1-x)p^2/m^2},\\
\end{eqnarray}


\end{document}